\documentclass{PoS}
\usepackage{subcaption}
\usepackage{lineno}
\title{Test beam results of a Cylindrical GEM detector for the BESIII experiment}

\ShortTitle{Test beam results of the BESIII new CGEM-IT}

\author{\speaker{G.~Mezzadri$^{b,i}$}, M.~Alexeev$^f$, A.~Amoroso$^{f,l}$, R.~Baldini Ferroli$^{a,c}$, M.~Bertani$^c$, D.~Bettoni$^b$, F.~Bianchi$^{f,l}$, A.~Calcaterra$^c$, N.~Canale$^b$, M.~Capodiferro$^{c,e}$, V.~Carassiti$^b$, S.~Cerioni$^c$, JY.~Chai$^{a,f,h}$, S.~Chiozzi$^b$, G.~Cibinetto$^b$, F.~Cossio$^{f,h}$, A.~Cotta Ramusino$^b$, F.~De Mori$^{f,l}$, M.~Destefanis$^{f,l}$, J.~Dong$^c$, F.~Evangelisti$^b$, R.~Farinelli$^{b,i}$, L.~Fava$^f$, G.~Felici$^c$, E.~Fioravanti$^b$, I.~Garzia$^{b,i}$, M.~Gatta$^c$, M.~Greco$^{f,l}$, L.~Lavezzi$^{a,f}$, CY.~Leng$^{a,f,h}$, H.~Li$^{a,f}$, M.~Maggiora$^{f,l}$, R.~Malaguti$^b$, A.~Mangoni$^{d,k}$, S.~Marcello$^{f,l}$, M.~Melchiorri$^b$, M.~Mignone$^f$, G.~Morello$^c$, S.~Pacetti$^{d,k}$, P.~Patteri$^c$, J.~Pellegrino$^{f,l}$, A.~Pelosi$^{c,e}$, A.~Rivetti$^f$, M.~D.~Rolo$^f$, M.~Savri\'e$^{b,i}$, M.~Scodeggio$^{b,i}$, E.~Soldani$^c$, S.~Sosio$^{f,l}$, S.~Spataro$^{f,l}$, E.~Tskhadadze$^{c,g}$, S.~Verma$^i$, R.~Wheadon$^f$, L.~Yan$^f$ \\

 \rlap{$^a$ Institute of High Energy Physics, Beijing, China}\\
    \rlap{$^b$ INFN, Sezione di Ferrara, Italy}\\
  \rlap{$^c$ INFN, Laboratori Nazionali di Frascati, Italy}\\
  \rlap{$^d$ INFN, Sezione di Perugia, Italy}\\
  \rlap{$^e$ INFN, Sezione di Roma, Italy}\\
  \rlap{$^f$ INFN, Sezione di Torino, Italy}\\
  \rlap{$^g$ Joint Institute for Nuclear Research (JINR), Dubna, Russia}\\
  \rlap{$^h$ Torino Politecnico, Italy}\\
  \rlap{$^i$ University of Ferrara, Italy}\\
  \rlap{$^k$ University of Perugia, Italy}\\
  \rlap{$^l$ University of Torino, Italy}\\

E-mail: \email{gmezzadr@fe.infn.it}}

\abstract{
Gas detector are very light instrument used in high energy physics to measure the particle properties: position and momentum.
Through high electric field is possible to use the Gas Electron Multiplier (GEM) technology to detect the charged particles and to exploit their properties to construct a large area detector, such as the new IT for BESIII. The state of the art in the GEM production allows to create very large area GEM foils (up to 50x100 $\mathrm{cm}^2$) and thanks to the small thickness of these foils is it possible to shape it to the desired form: a Cylindrical Gas Electron Multiplier (CGEM) is then proposed.

The innovative construction technique based on Rohacell, a PMI foam, will give solidity to cathode and anode with a very low impact on material budget. The entire detector is sustained by Permaglass rings glued at the edges. These rings are used to assembly the CGEM, together with a dedicated Vertical Insertion System and moreover they host the On-Detector electronic. The anode has been improved w.r.t. the state of the art through a jagged readout that minimize the inter-strip capacitance.

The mechanical challenge of this detector requires a precision of the entire geometry within few hundreds of microns in the whole area.
In this contribution an overview of the construction technique, the validation of this technique through the realization of a CGEM, and its first tests will be presented.
These activities are performed within the framework of the BESIIICGEM Project (645664), funded by the European Commission in the action H2020-RISE-MSCA-2014.
}

\FullConference{5th International Conference on Micro-Pattern Gas Detectors (MPGD2017)\\
		22-26 May, 2017\\
		Philadelphia, USA}

\begin{document}

\section{Introduction}
The possibility to employ gas detector, especially Micro Pattern Gas Detector, as tracking elements is well-exploited. Due to their lightness and robustness, Gas Electron Multipliers (GEMs) \cite{GEM} are extremely favored for these types of applications.

BESIII will install a new Inner Tracker based on the Cylindrical GEMs in the summer of 2019. The CGEM-IT project is led by the Italian collaboration, but it involves Uppsala and Mainz Universities, INFN and the Institute of High Energy Physics (IHEP) of Beijing. In order to show that the GEM technology can reach the experimental requirements, several test beams were performed.

\section{BESIII and BEPCII}
\begin{figure}
\centering
\includegraphics[scale=0.18]{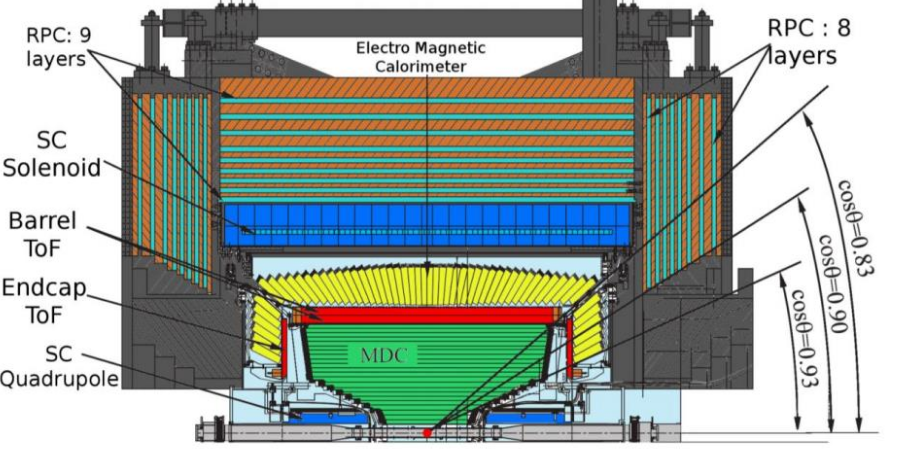}
\caption{Schematic view of the BESIII detector.}\label{detector}
\end{figure}

Beijing Spectrometer III (BESIII) is hosted at the Beijing Electron Positron Collider II (BEPC-II), at the IHEP. BEPC-II is a major upgrade of the preexisting accelerating structure. The beam energy spans from 1 GeV to 2.3 GeV, allowing BESIII to collect data in the energy regime from $2\, \mathrm{GeV}$ to $4.6\,\mathrm{GeV}$, which is frequently called $\tau-charm$ regime. BESIII is a central symmetry detector optimized for flavor physics. A Main Drift Chamber (MDC), combined with the 1 T solenoidal magnetic field, currently allows to reconstruct the position and momentum of the charged particles. It is composed of two parts and the innermost layers can be extracted in case of radiation damage. Time-of-Flight (TOF) detectors allow to reconstruct the time of passage of particles, to start the identification process. A $\mathrm{CsI(Tl)}$ electromagnetic calorimeter (EMC) measures the energy of neutral and charged particles. Resistive Plate Chambers (RPCs) are placed in the return yoke of the magnet to operate as Muon Counters. A schematic view of the detector is shown in Fig.~\ref{detector}. The detector covers $93\%$ of the $4\pi$ solid angle and it is divided in two parts: the barrel ($\vert \cos \theta < 0.82 \vert$) and the endcaps ($0.86 < \vert \cos \theta \vert < 0.93$). More details on the BESIII detector can be found in Ref.~\cite{BESIII_nim}.

\subsection{MDC Aging problems}
The eight innermost layers of the MDC are showing aging effects: in order to prevent large discharges, that can compromise the detector operations, the first layers are operating with voltages lower than nominal ones, affecting the reconstruction efficiency. Since the BESIII data taking has been extended up to 2028 and in order to improve the performance a new Inner Tracker based on CGEM technology will be installed in summer 2019. 

\section{The CGEM-IT project}
The CGEM-IT project will deploy a series of innovations and peculiarities in order to cope with the requirements provided by the BESIII collaboration, listed in Tab.~\ref{tab:requirements}.

\begin{table}[!h]
\begin{small}
\begin{center}
\centering
\begin{tabular}{|c|c|}
\hline 
\rule[-1ex]{0pt}{2.5ex} Value & Requirements \\ 
\hline 
\rule[-1ex]{0pt}{2.5ex} $\sigma_{xy}$ & $ \mathrm{\leq 130\, \mu m}$ \\ 
\hline 
\rule[-1ex]{0pt}{2.5ex} $\sigma_{z}$ & $\mathrm{\leq 1\, mm}$ \\ 
\hline 
\rule[-1ex]{0pt}{2.5ex} $\mathrm{dp/p}$ @ $1 \, \mathrm{GeV/c}$ & $0.5\%$ \\ 
\hline 
\rule[-1ex]{0pt}{2.5ex} Material Budget & $\mathrm{X_0 \leq 1.5\%}$ \\ 
\hline 
\rule[-1ex]{0pt}{2.5ex} Angular Coverage & $\mathrm{93\%}$ of $\mathrm{4\pi}$ \\ 
\hline 
\rule[-1ex]{0pt}{2.5ex} Particle Rate & $\mathrm{\sim 10^4 \, Hz/cm^2}$ \\ 
\hline 
\rule[-1ex]{0pt}{2.5ex} Minimum Radius & 65.5 $\mathrm{mm}$ \\ 
\hline 
\rule[-1ex]{0pt}{2.5ex} Maximum Radius & 180.7 $\mathrm{mm}$ \\ 
\hline 
\end{tabular}
\end{center}
\end{small}
\caption{List of the requirements provided by the BESIII collaboration for the new Inner Tracker}\label{tab:requirements}
\end{table}

The CGEM-IT will consist of 3 layers of triple GEM detectors, shaped as cylinders. The mechanical rigidity is provided by gluing the GEM foils around moulds of fixed radii. Permaglass rings are used to operate as gas sealing structure and gap spacers.  A sandwich of a PMI foam, called Rohacell, and kapton is used in order to provide mechanical rigidity to anode and cathode electrodes. Rohacell is a very light material that limits the material budget to $0.3\, \mathrm{X_0}$ per layer. A \textit{jagged} anode is used to reduce the inter-strip capacitance up to $30\%$ \cite{isabella}.

The most challenging requirement for the CGEM-IT development is the spatial resolution of $\sigma_{xy} < 130 \, \mathrm{\mu m}$ in 1 Tesla magnetic field. A dedicated ASIC is being developed to provide time and charge information for each strip. In order to assess that the CGEM-IT can reach the required performance an extensive series of test beams was performed in the last few years within the test beam activities of the RD51 Collaboration of CERN.

\section{Test Beam results}
\begin{figure}
\begin{center}
\includegraphics[scale=0.5]{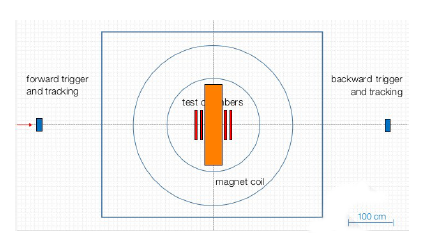}
\caption{Sketch of the setup for the test beams. Here is the one used for the cylinder, but the one with planar chambers have similar features.}\label{test_setup}
\end{center}
\end{figure}

The tests were performed both on $10 \times 10 \, \mathrm{cm}^2$ planar GEM chambers and on a cylindrical prototype of the dimension of the second layer of the final CGEM-IT.
All the tests were performed in H4 line of the SPS, in CERN North Area. Since the CGEM-IT will operate in a magnetic field, all the test chambers were placed inside Goliath, a dipole magnet that can reach up to 1.5 T in both polarities. Pion and muon beams were used: the momentum of the beam is 150 GeV/c. Scintillators were placed upstream and downstream the magnet to operate as trigger. The typical used setup  is sketched in Fig.~\ref{test_setup}, where the setup with the cylindrical prototype is shown.

\subsection{Position reconstruction with Charge Centroid}
The Charge Centroid (CC) readout uses of the charge induced on each strip to reconstruct the position where the particle passed. The $x_{CC}$ is defined has followed:
\begin{equation}
x_{CC} =  \frac{\sum_i x_i q_i}{\sum_i q_i}
\end{equation}

where $x_i$ is the i-th strip coordinate and $q_i$ is the charge induced on the i-th strip. If the avalanche has a Gaussian shape at the anode, the CC works properly and improves the best achievable digital readout resolution, that is of the order of $p/\sqrt{12}$, where $p$ is the strip pitch ($p = 650\, \mathrm{\mu m}$. 

\subsubsection{Results with no magnetic field applied}
We tested planar and cylindrical GEMs spatial resolution and efficiency with no magnetic field applied.

\begin{figure}
\centering
\includegraphics[scale=0.22]{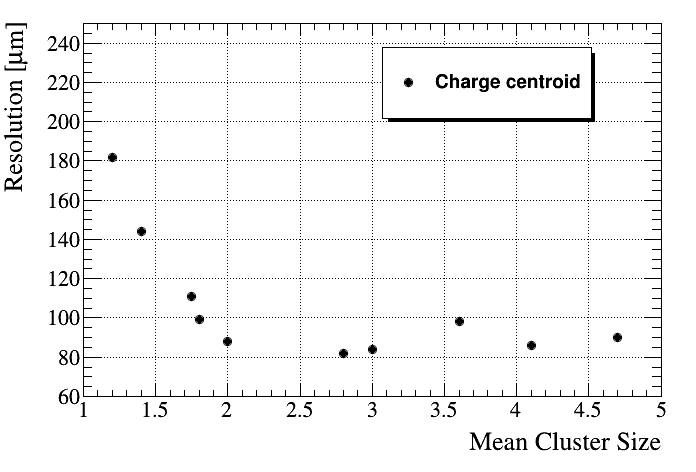}
\caption{Spatial resolution with respect to the cluster size.}\label{res_noB}
\end{figure}

\paragraph{Planar GEMs} The analysis of the data shows that an efficiency plateau can be reached starting from total gains of about 5000 with a typical value around $98\%$. In Fig.~\ref{res_noB} the resolution with respect to the mean cluster size is plotted.  It is possible to clearly notice that a spatial resolution below $100 \, \mathrm{\mu m}$ can be achieved for very different cluster sizes. An extensive collection of results for planar GEMs with no magnetic field can be found in Refs. \cite{results,results_1}.

\paragraph{Cylindrical prototype}
We have tested a cylindrical prototype. In order to simplify the reconstruction, only the $\phi$ view was instrumented. The goals were to test the construction procedure and to have a first comparison between a cylindrical and a planar GEM.

\begin{figure}[!h]
\centering
\begin{subfigure}[!h]{0.4\textwidth}
\includegraphics[width=\textwidth]{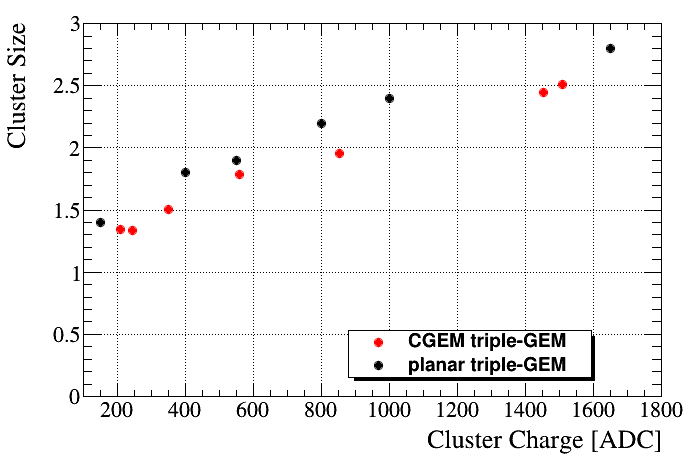}
\caption{Cluster size with respect to the charge in ADC units.}\label{cl_size_cyl}
\end{subfigure}
\quad
\begin{subfigure}[!h]{0.4\textwidth}
\includegraphics[width=\textwidth]{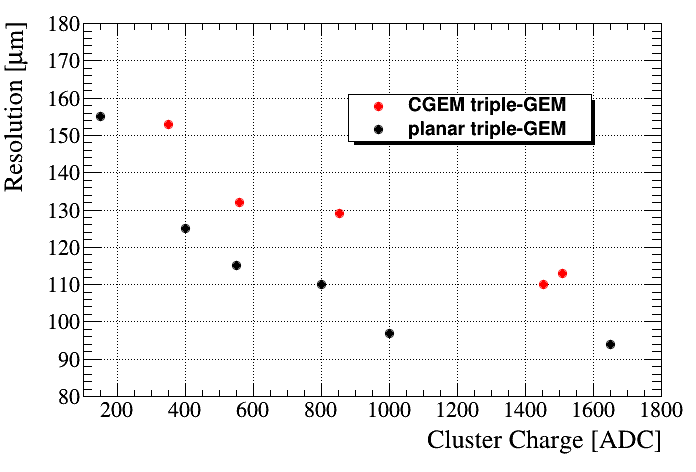}
\caption{Resolution with respect to the charge in ADC units}\label{res_cyl}
\end{subfigure}
\caption{Study of the performance of planar (red circles) and cylindrical (black circles) prototypes.}
\end{figure}

The cluster size and the spatial resolution are investigated as a function of the cluster charge. The resolution is measured as the width of the distribution of the differences between the reconstructed position of the front and the back of the cylinder. Fig.~\ref{cl_size_cyl} and Fig.~\ref{res_cyl} show the summary plots for cluster size and resolution respectively.

CGEM provide similar response to the planar GEMs. Further investigations are on-going in order to deeply understand the CGEM performance.  

\subsubsection{Results with magnetic field}
\begin{figure}[!h]
\centering
\begin{subfigure}[t]{0.38\textwidth}
\includegraphics[width=\textwidth]{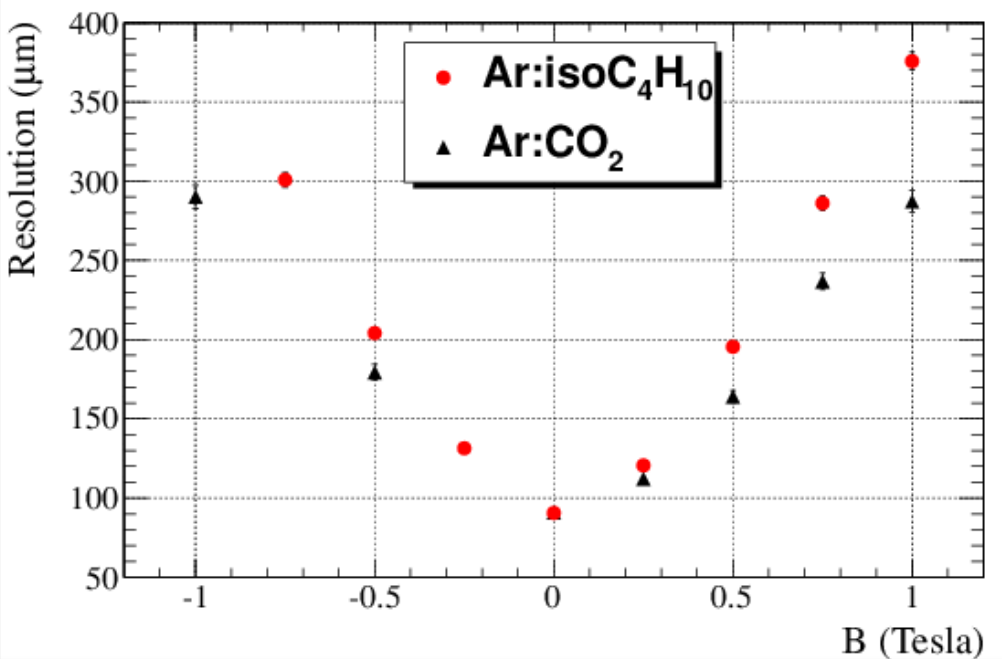}
\caption{Resolution with respect to the magnetic field strength for $\mathrm{Ar/iC_4H_{10}}\, (90/10)$ (red circles) and $\mathrm{Ar/CO_2}\,(70/30)$ (black triangles).}\label{res_B}
\end{subfigure}
\quad
\begin{subfigure}[t]{0.4\textwidth}
\includegraphics[width=\textwidth]{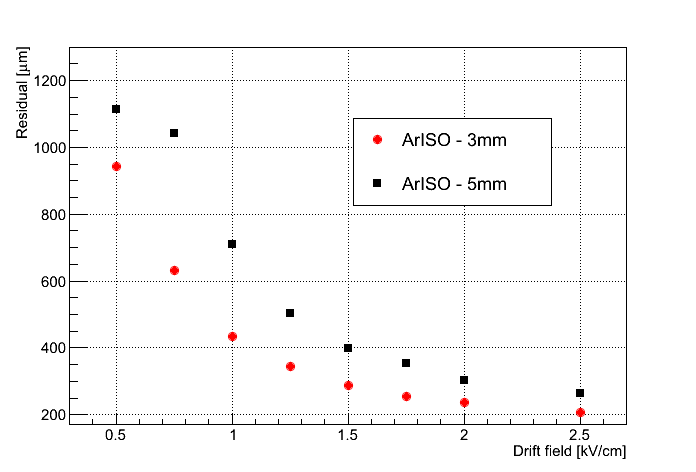}
\caption{Resolution with respect to drift field in 1 T magnetic field for two different drift gaps. 3 mm drift gap (red circles), 5 mm drift gap (black triangles).}\label{res_vs_drift}
\end{subfigure}
\caption{Studies of the resolution in presence of a magnetic field.}
\end{figure}

Planar GEMs performance in magnetic field was studied. The presence of an external magnetic field induces a deformation of the avalanche shape at the anode due to the Lorentz force: the charge centroid method performance degrades almost linearly with the magnetic field as shown in Fig.~\ref{res_B}.

It is still possible improve the performance by a proper optimization of the drift field as shown in Fig.~\ref{res_vs_drift}. With the proper choice of gas mixture ($\mathrm{Ar/iC_4H_{10}}\, (90/10)$) and drift field (2.5 kV/cm) is possible to achieve the unprecedented resolution of $\sigma_x ~ 190 \, \mathrm{\mu m}$ in 1 Tesla magnetic field.

\subsubsection{Combination of non-orthogonal incident tracks and magnetic field}
\begin{figure}
\begin{center}
\includegraphics[scale=0.35]{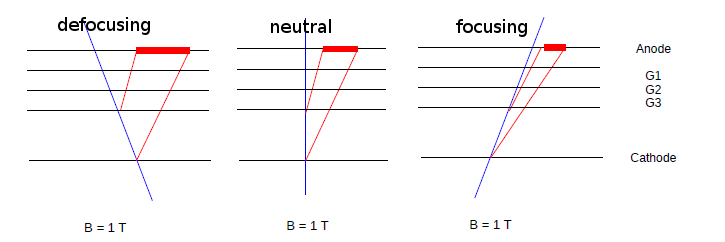}
\caption{Effects of the combination of magnetic field and incident tracks angles. (Left) $\alpha_{Lorentz} \neq \alpha_{track}$, $\alpha_{track} \neq 0$, defocusing. (Center) $\alpha_{Lorentz} \neq \alpha_{track}$, $\alpha_{track} = 0$, standard. (Right) $\alpha_{Lorentz} \sim \alpha_{track}$, focusing.}\label{focusing}
\end{center}
\end{figure}

In the final BESIII environment, the CGEM-IT will have to reconstruct tracks with very different incident angles in 1 Tesla magnetic field. Fig.~\ref{focusing} shows the possible combinations of effects. Two angles appear, one due to the Lorentz force ($\alpha_{Lorentz}$) and the other due to the non-orthogonal incident angle ($\alpha_{track}$). Different combinations of the two angles produce different charge shapes at the anode, and thus the resolution of CC degrades or improves accordingly. The CC gives the best performance when $\alpha_{Lorentz} \sim \alpha_{tracks}$ (right plot on the sketch)

\subsection{Position reconstruction with $\mu$TPC readout}
$\mu$TPC readout is an innovative  method that exploits the few millimeters drift gap as Time Projection Chamber. Indeed, the time of arrival of the induced charge on the strip can be used to reconstruct the first ionization position in the drift gap and thus improve the spatial resolution. To fully operate in $\mu$TPC mode, it is necessary to know both th drift velocity and the time of arrival of the induced signals. The latter is extracted from the electronics, while the former can be extracted from Garfield simulations. 

Once the different first ionization positions are reconstructed in the drift gap, it is possible to compute the track position by means of a linear fit to the found bidimensional points. The position is extracted by the following equation:

\begin{equation}
x_{\mu TPC} = \frac{\frac{gap}{2} - b}{a}
\end{equation}

where $a$ and $b$ are the straight line parameters and $gap$ is the drift gap thickness. This innovative readout was firstly developed for ATLAS MicroMEGAS \cite{ATLAS_utpc}. This is the first application of the method to GEMs in magnetic field. All the following results are extracted only from planar GEMs.

%
%

\subsubsection{Results with non-orthogonal tracks in 1 T magnetic field}
\begin{figure}
\begin{center}
\includegraphics[scale=0.26]{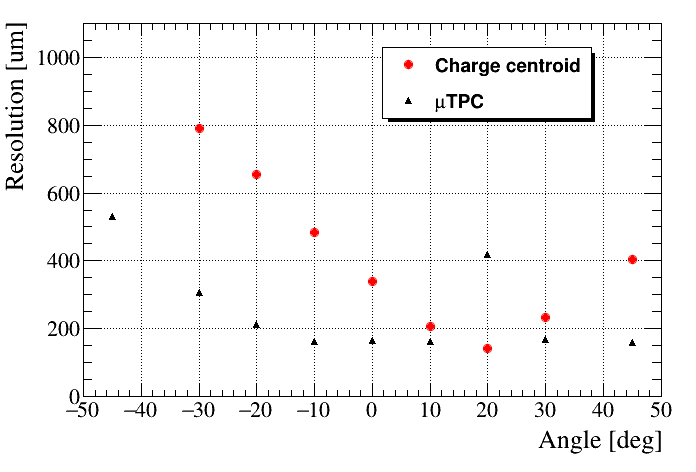}
\caption{Resolution with respect to the incident angle of the track in 1 Tesla magnetic field.}\label{utpc_angles1T}
\end{center}
\end{figure}

The presence of the magnetic field produces an interplay between the Lorentz angle and the incident angle to determine the final shape of the avalanche, and thus, to the performance of the $\mu$TPC readout. The Lorentz angle is determined by the gas mixture, the drift fields and the external magnetic field applied, while the incident tracks angle is varied by rotating the chambers. Fig.~\ref{utpc_angles1T} shows the spatial resolution with respect to the track incident angle for $\mathrm{Ar/CO_2}\, (70/30)$ gas mixture, with 1.5 kV/cm drift field and 1 Tesla magnetic field with respect to the incident track angle. The red circles and the black triangles represent the planar and the cylindrical prototype respectively.

A spatial resolution better than $200\, \mathrm{\mu m}$ has been achieved for a very wide set of incident angles with the $\mu$TPC mode. The focusing effect is already present at $10^\circ$, but has maximum impact when $\alpha_{track} \sim \alpha_{Lorentz} \simeq 20^\circ$ . Here it affects the result for the $\mu$TPC, but, as expected, the charge centroid method has its best resolution. A merge of the two techniques will combine the $\mu$TPC and CC information in order to achieve a better resolution and match the BESIII requirements.

\section{Final remarks}
A series of test beams has been performed to validate the cylindrical GEM technology and to address its limits in terms of spatial resolution in high magnetic field. 

The first test beam with the cylindrical GEM shows that the CGEM technology can operate in beam condition with performance similar to the planar one. More indications will come from the analysis of the data from the  test beam performed in July 2017, with the actual layer 1 (the innermost in the final BESIII CGEM-IT) and from extensive cosmic rays data taking.

In order to satisfy the spatial resolution requirement, two readout modes are deployed. With the conventional Charge Centroid method it is possible to achieve the unprecedented resolution of $\sigma_{x} \sim 190 \, \mathrm{\mu m}$ with orthogonal tracks in magnetic field. The $\mu$-TPC readout mode improves and overcomes the charge centroid limits granting a spatial resolution lower than $200\, \mathrm{ \mu m}$ for a large angle interval. These results represent the world best results for GEM in magnetic field. By a merge of the two readout modes will be possible to satisfy the BESIII requirements. 


\end{document}